\input harvmac

\def\vep{\varepsilon}

\def\vp{{\bf p}}

\def\be{\begin{equation}}
\def\ee{\end{equation}}

\def \k {\kappa} 

\def \g {\gamma}
\def \del {\partial}
\def \bd {\bar \partial }

\def \ha{{\textstyle{1\over 2}}}

\def \a {\alpha}
\def \b {\beta}

\def \s {\sigma}

\def \m {\mu}
\def \n {\nu}
\def \vp {\varphi }
\def \l {\lambda}
\def \t {\theta}
\def \td {\tilde }

\def \sm {$\s$-model }

\def \P {\Phi}

\def \inv {^{-1}}
\def \ov {\over }
\def \four{{\textstyle{1\over 4}}}

\def \sin{{\rm sin}}
\def \cos{{\rm cos}}
\def\[{\left [}
\def\]{\right ]}
\def\({\left (}
\def\){\right )}

\def \lr { \lref}
\def\np {{  Nucl. Phys. }}
\def \pl {{  Phys. Lett. }}
\def \mpl {{ Mod. Phys. Lett. }}
\def \prl {{  Phys. Rev. Lett. }}
\def \pr  {{ Phys. Rev. }}

\def \cqg {{ Class. Quant. Grav. }}

\lr \duh {A. Dabholkar, G.W. Gibbons, J. Harvey and F. Ruiz Ruiz,  \np
B340 (1990) 33;
A. Dabholkar and  J. Harvey,  \prl
63 (1989) 478.
}
\lr\sussk{L. Susskind, hep-th/9309145.}
\lr\mon{J.P. Gauntlett, J.A. Harvey and J.T. Liu, \np B409 (1993) 363.}
\lr\chs{C.G. Callan, J.A. Harvey and A. Strominger, 
\np { B359 } (1991)  611; in {\it 
Proceedings of the 1991 Trieste Spring School on String Theory and
Quantum Gravity}, J.A. Harvey {\it et al.,}  eds. (World Scientific, 
Singapore
1992).}

\lr\MV {J.C. Breckenridge, R.C. Myers, A.W. Peet  and C. Vafa, HUTP-96-A005,  hep-th/9602065.}
\lr\ssk{E. Halyo, A. Rajaraman and L. Susskind, hep-th/9605112. }

\lr \peet {A.W.  Peet, \np  {B456} (1995) 732.} 

\lr \US{M. Cveti\v c and  A.A.  Tseytlin, 
\pl { B366} (1996) 95, hep-th/9510097. 
}
\lr \USS{M. Cveti\v c and  A.A.  Tseytlin, 
\pr D53 (1996) 5619;
A.A. Tseytlin, \mpl A11 (1996) 689.
}
\lr\LW{ F. Larsen  and F. Wilczek, 
 hep-th/9511064;   hep-th/9604134.}
\lr\LWW{ F. Larsen  and F. Wilczek, 
    hep-th/9604134.}

\lr\TT{A.A. Tseytlin, \mpl A11 (1996) 689, hep-th/9601177.}
\lr \HT{ G.T. Horowitz and A.A. Tseytlin,  \pr { D51} (1995) 2896.}
\lr\khu{R. Khuri, \np B387 (1992) 315; \pl B294 (1992) 325.}
\lr\CY{M. Cveti\v c and D. Youm,
 UPR-0672-T, hep-th/9507090; UPR-0675-T, hep-th/9508058; 
  \pl { B359} (1995) 87, 
hep-th/9507160.}

\lr \CM{ C.G. Callan and  J.M.  Maldacena, 
  hep-th/9602043.} 
\lr\SV {A. Strominger and C. Vafa,   hep-th/9601029.}

\lr\HLM{ G. Horowitz, D. Lowe and J. Maldacena, hep-th/9603195. }

\lr\MV {J.C. Breckenridge, R.C. Myers, A.W. Peet  and C. Vafa, HUTP-96-A005,  hep-th/9602065.}
\lr\vijay{V. Balasubramanian and F. Larsen, hep-th/9604189.}
\lr \US{M. Cveti\v c and  A.A.  Tseytlin, 
\pl { B366} (1996) 95, hep-th/9510097. 
}
\lr\LW{ F. Larsen  and F. Wilczek, 
  hep-th/9511064.    }

\lr\DasMat{S. Das and S. Mathur, hep-th/9601152.}

\lr\ruts{J.G. Russo and A.A. Tseytlin, 
 \np B449 (1995) 91.}
\lr \dabo {A. Dabholkar,  \np B439 (1995) 650;
D.A. Lowe and A.  Strominger, \pr {D51} (1995) 1793.}

\lr\mon{J.P. Gauntlett, J.A. Harvey and J.T. Liu, \np B409 (1993) 363.}

\lr\mina{M.J. Duff, J.T. Liu and R. Minasian, 
\np B452 (1995) 261, hep-th/9506126.}
\lr\dvv{R. Dijkgraaf, E. Verlinde and H. Verlinde, hep-th/9603126;
hep-th/9604055.}
\lr\gibb{G.W. Gibbons and P.K. Townsend, \prl  71
(1993) 3754, hep-th/9307049.}
\lr\town{P.K. Townsend, hep-th/9512062.}
\lr\kap{D. Kaplan and J. Michelson, hep-th/9510053.}

\lr\ast{A. Strominger, hep-th/9512059.}
\lr \ttt{P.K. Townsend, hep-th/9512062.}
\lr \papd{G. Papadopoulos and P.K. Townsend, hep-th/9603087.}
\lr\jch {J. Polchinski, S. Chaudhuri and C.V. Johnson, 
hep-th/9602052.}

\lr \gig{G.W. Gibbons, M.J. Green and M.J. Perry, 
hep-th/9511080.}

\lr \CMP{ C.G. Callan, J.M.  Maldacena  and A.W. Peet, 
PUPT-1565,  hep-th/9510134.  } 

\lr \dab{  A.~Dabholkar, J.~Gauntlett, J.~Harvey and D.~Waldram, 
 hep-th/9511053.   }
\lr\verli{
K. Schoutens, H. Verlinde and E. Verlinde, \pr D48 (1993) 2670.
}

\lr\malda{ J. Maldacena, hep-th/9605016.}

\lr \US{M. Cveti\v c and  A.A.  Tseytlin, 
\pl {B366} (1996) 95, hep-th/9510097.  
}
\lr\mast{J.M. Maldacena and A. Strominger, hep-th/9603060.}
\lr \CY{M. Cveti\v c and D. Youm,
 \pr D53 (1996) 584, hep-th/9507090.  }
 \lr\kall{R. Kallosh, A. Linde, T. Ort\' in, A. Peet and A. van Proeyen, \pr { D}46 (1992) 5278.} 
\lr \grop{R. Sorkin, Phys. Rev. Lett. { 51 } (1983) 87;
D. Gross and M. Perry, Nucl. Phys. { B226} (1983) 29. 
}
\lr \myers{C. Johnson, R. Khuri and R. Myers, hep-th/9603061.} 
\lr \rusus{J.G. Russo and L. Susskind, \np {B437} (1995) 611.}
\lr \myk{R.R. Khuri and R.C. Myers, hep-th/9512061.}

\lr \KT{I.R. Klebanov and A.A. Tseytlin, hep-th/9604089.}

\lr \CYY{M. Cveti\v c and D. Youm, unpublished.}
\lr \CS{M. Cveti\v c and  A. Sen, unpublished.}
\lr \AT{ A.A. Tseytlin, hep-th/9604035.}

\lr \green{M.B. Green and M. Gutperle, hep-th/9604091.}
\lr \sen{A. Sen, \mpl  { A10} (1995) 2081.} 
\lr \gib{G.W.  Gibbons and K. Maeda, \np {B298} (1988) 741.} 
\lr \dabb{  A. Dabholkar, J.P. Gauntlett, J.A. Harvey and D. Waldram, 
  hep-th/9511053.   }
\lr \CMP{ C.G. Callan, J.M.  Maldacena  and A.W. Peet, 
PUPT-1565,  hep-th/9510134.  } 

\lr\susm{J. Maldacena and L. Susskind, hep-th/9604042.}
\lr\DasMat{S. Das and S. Mathur, hep-th/9601152.}

\lr \duf{M.J.  Duff, S. Ferrara, R.R. Khuri and J. Rahmfeld, 
\pl {B356} 
(1995) 479, hep-th/9506057.}

\lr \lars { F. Larsen, private communication.}
\lr \TTT{A.A. Tseytlin,  hep-th/9603099.}
\lr \garf{D. Garfinkle, \pr D46 (1992)  4286.}

\lr \ssen{ 
A. Sen,  \np B388 (1992) 457; D. Waldram, \pr D47 (1993)  2528.}

\lr \dul{M.J.  Duff and J.X. Lu,  \np { B354}  (1991) 141.}

\lr \ttt{A.A. Tseytlin, \pl B363 (1995) 223, hep-th/9509050.}
\lr \TET{A.A. Tseytlin, \pl B251 (1990) 530.}
\lr \lowe{D.A. Lowe and A. Strominger, \prl {73} (1994) 1468, hep-th/9403186.}
\lr \HS {G. Horowitz and A. Strominger, 
hep-th/9602051.}
\lr \HMS {G. Horowitz, J.M. Maldacena  and A. Strominger, 
hep-th/9603109.}

\lr \KT{I.R. Klebanov and A.A. Tseytlin, hep-th/9604166.}
\lr\dull{M.J. Duff and J.X. Lu, \pl B273 (1991) 409. }
\lr\hos{G.T.~Horowitz and A.~Strominger, Nucl. Phys. { B360}
(1991) 197.}
\lr\ght{G.W. Gibbons, G.T. Horowitz and P.K. Townsend, \cqg 12 (1995) 297,
hep-th/9410073.}
\lr\CYYY{M. Cveti\v c and D. Youm, hep-th/9603100.}
\lr\rutse{J.G. Russo and A.A. Tseytlin, \np {B461} (1996) 131.}
\lr\DVV{R. Dijkgraaf, E. Verlinde and H. Verlinde, hep-th/9603126.}
\lr \kallosh {E. Bergshoeff, I. Entrop and R. Kallosh,
 \pr D49 (1994) 6663.}
\lr\HOMA{G. Horowitz and D. Marolf, hep-th/9605224.}
\lr\TSS{A.A. Tseytlin, hep-th/9605091.}

\baselineskip8pt
\Title{
\vbox
{\baselineskip 6pt{\hbox{  }}{\hbox
{CERN-TH/96-145}}{\hbox{hep-th/9606031}} {\hbox{
  }}} }
{\vbox{\centerline  {The string spectrum on the horizon    }
\vskip2pt 
\centerline        { of a non-extremal black hole }
\vskip4pt
}}
\vskip -20 true pt
\phantom {\refs{\SV, \HMS } }
\phantom {\refs{\sussk, \rusus, \sen , \peet , \CMP, \dab , \LW, \USS ,\TSS, \ssk }}

\centerline  { J.G. Russo
 }

 \smallskip \bigskip
 
\centerline{\it  Theory Division, CERN}
\smallskip

\centerline{\it  CH-1211  Geneva 23, Switzerland}

\bigskip\bigskip\bigskip\bigskip

\centerline {\bf Abstract}
\medskip
\baselineskip10pt
\noindent

We investigate the conformal string $\sigma $-model
corresponding to a general five-dimensional 
non-extremal black hole solution.
In the horizon region the theory reduces to an exactly solvable
conformal field theory.
We  determine  the modular invariant
 spectrum of physical string states, which 
expresses the Rindler momentum operator in terms of
three charges and string oscillators.
For  black holes with  winding and Kaluza-Klein  charges, 
we find that  states made with only right-moving excitations  
have ADM mass equal to the  black hole ADM mass, and thus they can be used 
as sources of the gravitational field.
A discussion on  statistical entropy is included.

\medskip
\Date {June 1996}
\noblackbox
\baselineskip 16pt plus 2pt minus 2pt
\newsec{Introduction}
Recently two complementary approaches  have shed light
on the statistical origin of the entropy of extremal
black holes  in string theory.
One 
is based on the D-brane representation of string solitons with
RR charge (see e.g. \refs{\SV , \HMS }). 
This permits the counting  of all BPS quantum states  
in the weakly coupled theory with given charges. This number is a topological
quantity and it is unchanged in extrapolating from weak coupling to
strong coupling where the black hole appears;  thus it represents
the degeneracy of the ground state of these extremal black holes.
This method has provided a number of concrete results, although the
extension beyond extremality (for black holes with Schwarzschild radii
much larger than the compactification scale)  appears to be problematic. 
The other approach, pursued for instance in refs. 
\refs {\sussk - \ssk },
 is based on the idea  of interpreting   the statistical 
  entropy of  black holes with NS charges as originating from   oscillating  states of a  string source.

This paper follows the line of the second approach. 
We will consider the five-dimensional solution studied in 
refs. \refs {\CYYY ,  \HMS },
describing a  family
of non-extremal black holes,
which includes the standard five-dimensional Reissner-Nordstr\"om solution.
In sect.~2 it will be shown that the 
$\sigma $-model becomes, in the vicinity of the event horizon,  an exactly  solvable conformal field theory, 
which  is  formally related to a particular member  of the class of models solved
in ref.~\ruts \ (or the supersymmetric version given in ref.~\rutse ).
In sect.~3 the corresponding (type II) superstring model will be solved and
  the exact physical spectrum will be determined. This
 has the expected $T$-duality
invariance under the exchange of winding and Kaluza-Klein 
momentum charges.
The simplicity of the horizon theory 
opens the possibility of having 
a tractable framework to quantitatively
investigate certain types of  interactions with infalling matter.

While for extremal black holes the derivation of the
statistical entropy reduces to the counting of
BPS states (i.e. the counting of all marginal supersymmetric 
deformations of the 
conformal $\s $-model describing black holes with the same
asymptotic charges), for
 non-extremal black holes 
it is unclear what states should be counted. 
The formalism developed  here might in principle be of use to derive a statistical entropy,
 under the
basic assumption   that the entropy of a non-extremal black hole
originates from non-supersymmetric perturbations of the conformal
$\s $-model (or non-BPS excitations of the string source).
In order to count all states with  given energy and  charges, 
one then needs to establish a correspondence between the string energy 
due to the oscillations and the black hole ADM mass. 
In sect.~4 we include some remarks on this point,
though  a complete treatment is outside the
scope of this paper.
An appendix gives further details of the solution to the horizon model.

Let us first describe the extremal black hole
recently investigated in \TSS .
The background is represented by the $D=10$  supersymmetric conformal \sm\
with the following bosonic part 
\refs{\HT }
  \eqn\lag{
L = (G_{\m\n} +   B_{\m\n })(x)  \del X^\m \bd X^\n  + {\cal R}\P (x) }
$$= 
 F(x)  \del u \left[\bd v +  K(x) \bd u  \right] + 
(g_{mn} +   B_{mn })(x)  \del x^m \bd x^n  + \del y_a \bd y_a 
+  {\cal R}\Phi   \ , $$
where  $u= y  +t, \ v=y  -t$, 
$x_m$ ($m=1,2,3,4$) are non-compact spatial coordinates, 
 $y_a$ ($a= 1,2,3,4$), $y $  are coordinates of the  5-torus, and 
($r^2\equiv x_m x_m $)  
\eqn\pep{    g_{mn} = f(x) \delta_{mn}\   ,  \ \ \ \   
H^{mnk} = - {2\ov \sqrt{ g} }  \epsilon ^{mnkp}\del_p \phi 
\ , \ \ e^{2\phi}=f \ ,\ \ \ 
e^{2\P} = Ff \ , 
}
\eqn\choi{
f= 1 +{P\ov r^2} \ , \ \ \ \ \ 
F\inv = 1 +{Q_1\ov r^2}\  , \ \ \ \ \ 
K={Q_2\ov r^2} \ . }
The ten-dimensional $B_{\mu\nu}$ has both electric $Q_1$ and magnetic $P$ charges.
$Q_2$ corresponds to adding a Kaluza-Klein momentum boost in the $y $ direction
($y \sim y +2\pi R$).
They are related to integers by ($\a'=1 $)
\eqn\nomb{
a Q_1= wR\ ,\ \ \ a Q_2= {m\ov R}\ ,\ \ \ \ a\equiv {\pi\ov 4G}\ ,
}
where  $G\equiv G_5$ is the five-dimensional Newton constant
(related to the volume $(2\pi)^4 V$ of the four-torus $a=1,2,3,4$,
and the radius $R$ by $G=\pi g^2/(4RV)$). 

Upon compactification along $y_1,...,y $ 
one finds the $D=5$ black hole (Einstein-frame) metric
\eqn\mee{
ds^2_5 =- \l^{-2/3} (r)  dt^2 +  \l ^{1/3}  (r)  (dr^2 + r^2 d\Omega^2_3) \ , 
}
$$
\l 
 = \( 1+ {Q_1\ov r^2}\) \( 1+ {Q_2\ov r^2}\) \( 1+ {P\ov r^2}\)\ . 
$$ 

In the throat region $r\to 0$  the theory is just described
by a direct product of $SL(2,R)$ ($t, y , r $) and SU(2)
(with Euler angles  associated with the angular coordinates
 of the three-sphere)
WZW models with equal levels proportional to $P$ \TSS .
It will be shown below that this conformal model does not describe the
extremal limit of the  horizon theory of a non-extremal black hole.

\newsec{ Conformal model for a non-extremal black hole }

\def\sa{r_0^2 {\rm sinh}^2\alpha }
\def\sg{r_0^2 {\rm sinh}^2\gamma }
\def\ss{r_0^2 {\rm sinh}^2\beta }

\def\e{\epsilon}

The non-extremal solution is conveniently parametrized in terms
of `boost' parameters \HMS , $\{ Q_1, Q_2, P, E \} \to \{\a,\g, \s, r_0\}$,
with the relations
\eqn\charges{
   Q_1 =   { r_0^2  \over 2 } \sinh 2 \a \ , \ \ \ 
  P =  { r_0^2\over 2} \sinh 2 \g \ , \ \ \ 
 Q_2 = {  r_0^2 \over 2 } \sinh 2 \b \ ,\ \ \ 
}
\eqn\mss{E=  {  a r_0^2  \over 2 } 
(\cosh 2 \alpha + \cosh 2 \gamma + \cosh 2 \b )\ ,\ \ \ \ a\equiv {\pi\ov 4G}\ .
}
In terms of the $\sigma $-model metric, the solution reads
\eqn\metric{
ds^2=\( 1+{\sa \ov r^2} \) ^{-1}\[ -dt^2+dy ^2 +{r_0^2\ov r^2}
(\cosh \b dt+ \sinh \b dy )^2\] +dy_a^2
}
$$
+\( 1+{\sg \ov r^2}\) 
 \left[
\(1-{r_0^2 \over r^2}\)^{-1} dr^2 + r^2 d \Omega_3^2 \right] \ , 
$$
\eqn\dil{ 
e^{2\Phi } = \(1+   { \sg \over r^2 }\)\(1 + {\sa\over r^2 }\)^{-1}\ ,
\ \ \ \ \ \ H= 2P \e_3\ ,
}
where $\e_3$ is the volume form on the unit three-sphere.
The extremal limit corresponds to the limit 
$ r_0 \rightarrow 0 $ with at least one of the boost parameters
$ \alpha, \beta , \gamma \rightarrow \pm \infty $ keeping
                the associated charges \charges \ fixed.
For the extremal solution there exists a renormalization scheme in which
the corresponding supersymmetric $\s $-model \lag\ is conformal to
 all orders in the $\a' $ expansion. 
The non-extremal black hole is not of the chiral null form and is
likely to receive $\a '$ corrections in all renormalization schemes.
However, for large black holes  $\a' $ corrections to the metric \metric\  can be neglected
in the horizon region. In particular, they would   affect terms in
 the entropy formula which are subleading in the macroscopic limit. 
Only the leading
part of the entropy is expected to be universal (i.e. the same for
type II or heterotic embeddings), and given by the
Bekenstein-Hawking formula.

The five-dimensional (Einstein-frame) metric  arising upon dimensional reduction has the symmetric form:
\eqn\solnfd{
ds_5^2 =  - h^{-2/3} \(1-{r_0^2 \over r^2}\) dt^2 +h^{1/3} 
\[\(1-{r_0^2 \over r^2}\)^{-1} dr^2 + r^2 d \Omega_3^2 \right] \ ,
}
where
\eqn\deff{ h= 
\(1+{\sa\over r^2} \)\(1+{\sg\over r^2} \)\(1+{\ss\over r^2} \)\ .
}
The background represents a solution of extended $D=5$ supergravity (with
$N=1$ supersymmetry in the extremal case).
 The solution is
manifestly invariant under permutations of the three boost parameters as
required by $U$-duality (arising as a subgroup of the
 global $E_6$ symmetry of $D=5$ supergravity).
The event horizon is  at $r=r_0$, and
$r=0$ represents
a regular inner horizon. 
The entropy and the Hawking temperature are given by
\eqn\entropy{
S = {A_5\over 4 G_5} =
2 \pi a  r_0^3   \cosh \alpha \cosh \beta \cosh \gamma  \ ,
}
\eqn\thwk{T= {1 \over 2 \pi  r_0 \cosh\alpha \cosh \beta \cosh \gamma }\  .
}

It is convenient to introduce a new coordinate $\hat r$
defined by
\def\td {\tilde }
\eqn\cooo{
\hat r= {r\ov r_0}+\sqrt{{r^2\ov r_0^2}-1 }\ . \ \ 
}
The  $r, \Omega_3$ part of the metric \metric\ takes the form:
\eqn\trtr{
ds^2= {1\ov \hat r^2} ( r^2 +\sg  )  (d\hat r^2 +\hat r^2 d\Omega_3^2)
= r_0^2 (\cosh^2\rho +\sinh^2 \g ) (d\rho^2 + d\Omega_3^2)
\ ,\ 
}
\eqn\cooo{
 r=r(\hat r)=\ha r_0 (\hat r+ \hat r ^{-1})=r_0\cosh \rho \ \ ,\ \ \ \ \ \ 
\rho\equiv\ln \hat r \ ,
}
\eqn\angular{
d\Omega_3^2=d\t ^2 + \sin^2\t \ d\varphi ^2 +\cos^2\theta\  d\psi ^2\ .
}
The background can then be represented by the $D=10$  supersymmetric conformal \sm , with the  bosonic part ($\hat r^2= x_m x_m$) :
  \eqn\lagrr{
L =  F(\hat r)  \[ - \del \hat t \bd \hat t + \del \hat y  \bd \hat y  
 +{r_0^2\ov r^2(\hat r)}\del \hat t \bd \hat t  \] +
 B_{y  t} (\del y  \bd t - \bd y  \del t)
}
$$
 + (g_{mn} +   B_{mn })(\hat r)  \del x^m \bd x^n  
 + \del y_a \bd y_a 
+  {\cal R}\Phi   \ , \ \ \ \ \ \ \ \ 
 $$
\eqn\defin{
\hat y  \equiv  \sinh \b \ t+ \cosh \b \ y  \ ,\ \ \ \  
\hat t \equiv  \cosh \beta \ t+ \sinh \beta \ y  \ ,
}
where the dilaton and $H^{mnk}$ were given in eq. \dil , 
  $g_{mn} = \hat f(x) \delta_{mn}$ , and $B_{y  t}= Q_1 F/r^2$,
with
\eqn\choirr{
\hat f= {1\ov \hat r^2} (r^2(\hat r) + \sg   )  \ , \ \ \ \ \ 
F\inv = 1 +{\sa \ov r^2(\hat r)}\   . }
This background can be embedded in both type II and  heterotic
superstring theories, and it can be physically interpreted
 as representing a bound state of a closed macroscopic string 
wrapped $w$ times around $y $ (with a momentum boost $m/R$), 
and a solitonic five-brane wrapped
$P$ times around the five-torus (the special cases   
$w=0$ or $P=0$ describe the geometry of a macroscopic string or  a  five-brane).

In the horizon region $\rho\to 0 $ the  $\sigma $-model action takes 
the form
\eqn\tet{
 I  =  {1 \ov \pi }\int d^2 \s  L_{\rho \to 0} 
=  I_1 + I_2  \ ,
}
\eqn\iii{
I_1={1 \ov \pi \cosh^2\a }\int d^2 \s  \(  
  \del \hat y   \bd \hat y  -\rho^2 
(\del \hat t +\tanh\a \del \hat y) (\bd \hat t -\tanh \a \bd \hat y ) \)  \ ,
}
\eqn\iiii{
I_2= 
{ 1\ov \pi  }\int d^2 \s  \left(
\k ^2 \del \rho \bd \rho 
+\del \t' \bd \t' +\del \varphi'\bd \varphi'
+\del \psi'\bd \psi' \right)\ , \ \ \ \ \ \kappa \equiv  r_0\cosh \g \ .
}
where  the free internal coordinates
and the dilaton term  have been omitted.
We have also rescaled the angular coordinates so that they have a standard
kinetic term.\foot{
For large $\k $ ($\k ^2\gg \a' $), the dynamics in the horizon region can
be locally 
described by flat coordinates. Expanding $\t= \t_0+{\t '\ov {k }}$,
one has $\k ^2 \del \t\del \t\to\del\t'\del \t'$, and $\sin^2\t\cong \sin^2\t_0$,
so that 
$\varphi' ={\k }\sin\t_0 \varphi \  ,\ \ \psi' ={\k }\cos \t_0 \psi $.
We also drop a total derivative term $\del \varphi ' \bd\psi' -\bd\varphi ' 
\del\psi '$, which is justified
in the large-$\k $ limit.}
The new  coordinates $\t ', \varphi', \psi'$  are periodic with period proportional
to ${\k }$. For large $\k $  their winding numbers can be ignored,
so we can treat these coordinates just as free uncompact coordinates.
In the extremal limit $\k ^2$ reduces to the magnetic charge $P$.
For the Reissner-Nordstr\"om  case, $\a=\b =\g $,   $\kappa $ is just
 the  Schwarzschild radius $R_s=r_0\cosh\a $ (see eq.~\solnfd ). 
In general, it will depend on  the three charges \charges\ 
and on the ADM energy \mss .

It is easy to check by direct calculation that the $T$-duality invariance
of the original $\s $-model \lagrr\ is preserved in the horizon
theory \tet , i.e. a $T$-duality transformation in the $y$ direction leads
to the same $\s $-model \tet\  with $\a $ and $\b $ exchanged.

The $\s $-model \lag\ describing  the extremal black hole solution has a different ($SL(2,R)\times SU(2) $ WZW)
structure near the horizon \TSS .
This cannot be recovered  by taking 
the extremal limit on eq. \tet .\foot{
In the limit $r_0\to 0$, one has $\rho\to\infty $, so the expansion
for small $\rho$  leading to \tet\ does not apply.
To take the extremal limit, it is useful 
to introduce a new coordinate $\td \rho $ which is regular as $r_0\to 0$, by the shift  
$\td  \rho =\rho + \ln r_0/2$ , so that $r=e^{\rho} +{r_0^2\ov 4} e^{-\rho }$ .
} 
 The horizon limit
$r\to r_0$ and the extremal limit $r_0\to 0$ do not commute, i.e.
the horizon theory of the extremal black hole is in no
limit  associated with the physics of the non-extremal black hole.

\newsec{ Spectrum in the horizon region }

The analysis of the  physical
spectrum of the model \tet\ is of interest, in particular,  
to establish a correspondence
between the asymptotic parameters characterizing the black hole 
geometry
and the  quantum numbers of a string moving in the 
vicinity of the horizon.\foot{
 Earlier studies in this direction
can be found in refs. \refs {\sussk , \ \dabo }, where the string spectrum in the presence of
cone singularities
(associated with Euclidean Rindler space at finite temperature)
is discussed.} 
Throughout we  restrict the discussion to the sector where the winding number 
and Kaluza-Klein momentum of the string state coincide with those
of the black hole geometry.
We will consider the type II superstring theory embedding
of the background discussed in the previous section.
To simplify the notation, during part of the derivation  the fermionic fields will be  omitted.
The only substantial difference with the purely bosonic case (apart from straightforward
addition of  fermion mode oscillators) is a normal ordering constant.

Let us introduce Minkowski world-sheet coordinates,
$\s_{\pm }=\tau \pm \s$.
The non-trivial part of the Lagrangian \tet\ is 
\eqn\lagra{
\pi  L=\del_+z\del_-z - \td\rho ^2 (\del_+ T +A \del_+ z )
(\del_- T- A\del_- z) +\del_+\td  \rho \del_- \td\rho \ ,
}
where
\eqn\defi{
 z\equiv 
{\hat y \ov \cosh \a } \ , \ \ \ T\equiv  {\hat t\ov \k \cosh\a }\ ,
\ \ \ \td\rho\equiv \k \rho \ ,\ \ \ A\equiv {\tanh\a \ov r_0\cosh\g }\ . 
}
The boundary conditions for the  fields $X^\pm \equiv \td \rho e^{\pm T}$ and $z$, as follows from
$y \sim y +2\pi w R$, are given by
\eqn\bbbb{
X^\pm (\s +\pi )= e^{\pm 2\pi \vep } X^\pm (\s )\ , \ \ \  \ \ \  
\vep={2\pi wm\ov S}={a r_0\sinh\a \sinh\b \ov \cosh\g }\ ,
}
\eqn\bbzz{
z(\s +\pi )= z(\s )+ 2\pi w \td R\ ,\ \ \ \ \ \td R=R {\cosh\b \ov\cosh \a }\ ,
}
where $S$ is the black hole entropy, eq. \entropy .
We will also need to  relate  the momentum $p_z$ to the momentum 
$p_y=m/R=a Q_2=ar_0^2 \sinh \b \cosh\b $. More generally, consider the canonical
momenta $\Pi_y, \Pi_z, \Pi_T, \Pi_t $ conjugate to $y,z,T,t $.
From eqs. \defin \ and \defi , we find
\eqn\uno{
 \Pi_y=\Pi_z {\cosh\b \ov \cosh\a} +
\Pi_T {\sinh \b\ov r_0\cosh\g \cosh\a }\ ,
}
\eqn\tres{
  \Pi_t=\Pi_z {\sinh\b\ov \cosh\a } +
\Pi_T {\cosh \b\ov r_0\cosh\g \cosh\a }\ ,
}
and the inverse relation
\eqn\dos{
 \Pi_T=r_0 \cosh\g \cosh\a ( \Pi_t \cosh\b  -\Pi_y \sinh \b) \ ,
}
\eqn\cuat{
  \Pi_z=\cosh\a (-\Pi_t \sinh\b +\Pi_y \cosh \b) \ .
}
Using eq. \uno , we obtain
\eqn\pizz{
p_z= \int_0^\pi d\s \Pi_z= {m\ov \td R} - BP_T=
 {m\ov \td R} \(1-{2\pi P_T\ov S}\)\ , 
}
$$
P_T\equiv \int_0^\pi d\s \Pi_T\ ,\ \ B\equiv {\tanh\b \ov r_0\cosh\g }\ .
$$
$P_T$ generates translations in  Rindler time.

The structure of the theory \lagra\ is that of an interacting  theory with non-trivial boundary conditions. Nevertheless, it can be solved
exactly. Indeed, the Lagrangian \lagra\ has the same form as a
 world-sheet Lagrangian related to the ($a=\sqrt{3}$) 
Kaluza-Klein Melvin Universe \rutse \ :
\eqn\melvin{ 
\pi  L_{\rm Melvin}=\del_+ y' \del_-y' 
+ \rho ^2 (\del_+ \varphi + A  \del_+ y' )
(\del_- \varphi - A \del_- y') +\del_+  \rho \del_- \rho \ ,
}
where $y'$ is a  boson compactified on a circle of radius $R'$,
and $\rho, \varphi $ are polar coordinates.
More precisely, the Kaluza-Klein magnetic flux tube Universe is described
by the Lagrangian which is $T$-dual to \melvin\ in the $y'$ direction
(with the addition of other free directions).
A general class of magnetic flux conformal models interpolating between the dilatonic ($a=1$) Melvin model and the Kaluza-Klein Melvin model
(and their $T$-duals) were  solved in ref. \rutse , with the
spectrum and the modular invariant partition function determined  \ (a larger
family was previously treated in \ruts\ for the bosonic string theory).

Instead of repeating the derivation of ref. \rutse ,
we will  settle the appropriate dictionary between the models \lagra\ and
\melvin \ (the direct solution of the model \lagra\ is outlined in the appendix). For this purpose, it is convenient to
go to Euclidean Rindler time, $T\to -iT$, $P_T\to i P_T$ (at the end we shall return to
the Minkowski-signature Rindler time).
The coordinate $\varphi $ can then be identified with $T$ (and $A\to i A$).
The operator $P_{T}$ will be identified with the
angular momentum operator $\hat J=\hat J_R +\hat J_L $ of the model \melvin \ 
which generates shifts in $\varphi $.

There are two differences between the theories \lagra\ and \melvin \ :
a) in \melvin , the field $\rho e^{i\varphi }$ is single-valued
(being the physical Cartesian uncompact coordinate),
whereas in the theory \lagra\ this field obeys the boundary condition
\bbbb ; b) in the magnetic flux tube model \melvin ,
 the momentum $p_{y'}$ is equal to $m'/R'$,
whereas the analogous $p_z$ is given by eq.~\pizz .

The spectrum of the model \melvin \ is not invariant under a $T$-duality transformation
in $y'$ (the partition function of course it is).
In fact, $T$-duality maps to a different $\s$-model, the Kaluza-Klein
Melvin theory.\foot{This follows from the standard $T$-duality rules;
the Kaluza-Klein Melvin model has a vector gauge field, which under $T$-duality
goes into the axial gauge field of \melvin .
Instead, the ($a=1$) dilatonic Melvin theory has a purely left gauge
 coupling;  the model is  self-dual, with a $T$-duality
invariant spectrum \rutse .
}
However,  the physical spectrum of the model \lagra\
must be invariant under $T$-duality:
 our original Lagrangian \tet\  is
self-dual under a $T$-duality transformation in the $y$-direction (with the change $(\a,\b )\to (\b,\a )$).
 Indeed, we will show that, by taking into account
the two differences (a) and (b) between the models pointed out above, the resulting
spectrum will enjoy the expected symmetry.

The Hamiltonian corresponding to the model \melvin\ is given by 
(see eq. (6.8) in ref.~\rutse\ with $q=0$ and $\beta $ instead of $A$)
\eqn\Hamil{
\hat H=\hat N_L +\hat N_R - \ha p_t^2+\ha p_\a^2 + \ha p_{y'}^2 + \ha (wR- A \hat J)^2
-\hat\g (\hat J_R-\hat J_L) \ ,
}
\eqn\gami{
\hat\g = A p_{y'}\ ,\ \ \ \hat J=\hat J_R+ \hat J_L \ ,
} 
where the index $\a $  stands for extra free  coordinates.\foot{
For simplicity, we shall consider only states with zero winding number and
momentum in the internal coordinates $a=1,2,3,4$, so that $p_\a^2$ only 
contains contributions from the variables $\t ' ,\ \varphi ', \ \psi '$~.}
The operators $\hat N_{R,L}=N_{R,L}-a$
contain the free superstring theory Regge intercepts 
$a^{\rm (R)}=0, \ a^{\rm (NS)}=1/2$, 
   e.g.   in the Ramond sector,
\eqn\nnn{
 N^{\rm (R)}_R= \sum_{n=1}^\infty n \big( b^{\dagger }_{n+}b_{n+}+ b^{\dagger
}_{n-}b_{n-}
+ b^{\dagger }_{n\a} b_{n\a}  +  d^*_nd_n+d _{- n}d_{-n}^* + d _{-n\a}
d_{n\a}\big) \ ,  }
$$ [b_n, b_k^{\dagger}]=\delta_{nk}=\{ d_n, d_k^*\} \ .
$$
The free creation and annihilation operators $b_n, b_n^{\dagger}$ and 
$d_n , d_n^*$  are associated  with the bosonic degrees of freedoms
represented by $\rho e^{\pm i\varphi }$ and the fermionic superpartners.
Here the light-cone gauge has been used to remove oscillator modes
corresponding to the degrees of freedom associated with $y'$ and $t$
(the corresponding term $-\del_+ t\del_- t$ was omitted in eq.~\melvin\ since
$t$ is decoupled).
In the Ramond sector, the operator $ \hat J_R$ is given by
\eqn\angulr{
{\hat  J}_R= - b^{\dagger }_0 b_0 -\ha    +\sum_{n=1}^\infty \big(  b^{\dagger
}_{n+}b_{n+} - b^{\dagger }_{n-}  b_{n-} \big)+\hat K_R  = P_{R}\ , }
$$
 \hat  K^{\rm (R)}_R=-[d_0^*,d_0]  +
  \sum_{n=1}^\infty (d_{n}^* d_{n}+d_{-n}  d_{-n}^*) \  .   
$$
The expression of $\hat J_L= P_{L}$ is similar,   with  a reversal of the
 sign of the orbital angular momentum terms.

The analogue of $\hat \g $ for  the model \lagra \ is $\hat \gamma =iA p_z $.
However, because of the twisted boundary condition \bbbb\ in $X^\pm $,
 the parameter $\hat \gamma $ will be shifted by $i\vep $, so
 it must be replaced by 
\eqn\cambio{
\hat \gamma \ \longrightarrow \ \td \g =\hat \gamma +i\vep =
{4\pi i mw \ov S}+ AB P_{T} \ .
}
The case $mw=0$, implying $\td \gamma =0$, is special and it will be discussed separately.

Taking into account these changes, we can now write down
the Virasoro conditions, determining the physical spectrum of
the Euclidean horizon theory:\foot{In the bosonic theory the Hamiltonian has an extra  
normal-ordering contribution 
$\ha \td \g ^2$.} 
\eqn\vira{
\hat H_{\rm Eucl}=\hat N_L +\hat N_R +\ha p_\a^2 + 
\ha \( {m\ov \td R} - i B P_{T} \)  ^2 + \ha \( w \td R- i A P_{T}\) ^2
}
$$
-\( {4\pi i mw \ov S}+AB P_{T} \)  (P_{R}-P_{L})=0 \ , \ \ \ \ \ \ 
P_T=P_R+P_L \ ,
$$
\eqn\soro{
\hat N_R- \hat N_L=mw \ ,
}
where $A$ and $B$ have been defined in eqs. \defi \ and \pizz .

Let us now formally return to Minkowski-signature Rindler time by
$ P_{R,L}\to -i P_{R,L} $.  We obtain
\eqn\mmma{
\hat H=\hat L_0+ \hat {\td L_0}=\hat N_L +\hat N_R +\ha p_\a^2 + 
\ha \( {m\ov \td R} - B P_T \)  ^2 + \ha \( w \td R - A P_T\) ^2
}
$$
+AB (P_R^2-P_L^2)- {4\pi wm\ov S} (P_R-P_L)=0
$$
\eqn\soro{
\hat L_0 - \hat {\td L_0}=\hat N_R- \hat N_L-mw=0\ ,
}
Using the definition of $A, B$ and the relation between
the $\a,\b $ and $m, w$ (se eqs. \nomb , \charges ), eq.~\mmma\ can be written in the form
\eqn\spec{
2\hat H=2\hat N_L +2\hat N_R + p_\a^2 + 
\( {m^2\ov \td R^2 } + w^2 \td R^2 \) \(1- {2\pi P_T\ov S} \) ^2  
}
$$
-{8\pi  mw \ov S^2}(P_R-P_L)( S -\pi P_T ) =0 \ .
$$
The spectrum is thus invariant under 
the duality transformation
$\a \leftrightarrow \b $ (or $m \leftrightarrow w $, $R\leftrightarrow 1/R$).

In contrast with the Euclidean theory,  now the operators
$P_L$, $P_R$ are no longer angular momentum operators.
They are defined by 
\eqn\momr{
P_{L,R}=\int_0^\pi d\s \ \Pi_{L,R}\ ,\ \ \ \ 
\Pi _{L,R}={\td \rho^2 \ov 2\pi }  (\del _\pm T \pm A \del _\pm z)\ ,
\ \ \ P_T=P_R+P_L\ .
}
In terms of oscillator modes, they can still be {\it formally} written  
as in eq. \angulr\ with a factor $-i$ in front, but  it is convenient to return  to the conventional notation $\a_n^\mu $. This is done by defining
\eqn\alfa{
b_{n\pm }\equiv {\a_n^1 \pm \a_n^0 \ov \sqrt{2n} }\ , \ \ \ 
b_{n\pm}^\dagger \equiv {\a_{-n}^1 \mp \a_{-n}^0 \ov \sqrt{2n} }\ , \ \ \ \  n=1,2,... \ ,
}
 and similarly for the left modes $\td \a_n^\mu$.
The correct conjugation properties of the $b_n, b_n^\dagger $ operators 
follow from $(\a_{-n}^\mu )^\dagger = \a_n^\mu $ (see appendix). The resulting expressions for $\hat N_{R} $ and $P_R$ contain the familiar
contributions $-\a_{-n} ^0 \a_n ^0+ \a_{-n} ^1 \a_n ^1 $ and  
${i\ov n} (\a_{-n}^1 \a_n^0 - \a_{-n}^0 \a_n^1 )$. 
The way the time enters into the Lagrangian of the model \lagra\ 
is different from the analogue model \melvin . This makes a difference in the choice of
the light-cone gauge (for the Euclidean theory it makes of course  no difference).
In order to avoid negative-norm states in the spectrum, now the light-cone gauge must be used to remove the oscillators
$\a_n^0,\a_n^1 $, $n\neq 0$. As a result, only a constant zero mode contribution
$x^\pm ,\td x^\pm $ (related to $b_0, b_0^\dagger$ by 
$b_0, b_0^\dagger =\sqrt{-i }\ x^\mp $, 
$\td b_0, \td b_0^\dagger =\sqrt{i }\ \td x^\mp $)
remains in the Rindler operators $P_{R,L}$. Now the different operators 
 take the form ($mw \neq 0$)
\eqn\pppr{
P_R=x^+x^- \ ,\ \ \   P_L =\td x^+ \td x^-\  ,
\ \ \ \ \ \ [x^-,x^+]=i=[\td x^+ ,\td x^- ]\ ,
} 
\eqn\otro{
 N^{\rm (R)}_R= \sum_{n=1}^\infty  \big[ \a_{-nz}\a_{nz}
+ \a _{-n\a} \a _{n\a}  +  n(d_{-nz}d_{nz} + d _{-n\a}
d_{n\a}) \big] \ , 
}
$$
[\a_n^\mu , \a_k^\nu]=n\delta_{n+k}\eta ^{\mu\nu}\ ,\ \ \ \ \{ d_n, d_k\} =\delta_{n+k} \ ,
$$
where the modes with subindex $z$ are those associated with the field $z$ and its
superpartners.
The physical space is simply constructed by applying 
the free creation operators on the vacuum Fock state.
The full treatment of the zero mode sector $x^\pm ,\td x^\pm $ involves
long technical details, which we plan to present elsewhere.\foot{
 A system with a similar zero mode structure is discussed in ref.~\verli .} 
The calculation given here will not depend on the explicit form of
 $P_L , P_R$.

In the special case $mw=0=\td \g$, the fields $X^\pm $ are single-valued. As a result,
translational invariance of the  Rindler Cartesian plane 
$X^1=\td \rho\cosh T, \ X^0=\td\rho \sinh T $ is restored.
The zero-mode operators $ x^\pm $ and 
$\td x^\pm $, are traded by center-of-mass coordinates
and momentum $x_0, x_1$ and  $p_0, p_1$. The bosonic zero mode parts of
$P_L, P_R$ are replaced by $\ha (x_0 p_1-x_1 p_0)$,
and the Hamiltonian reads ($w=0$)
\eqn\moma{
\hat H= \hat N_R +\hat N_L +\ha p_\a^2 +\ha  p_1^2-\ha p_0^2+
 {m^2\ov 2\td R^2 }  \(1- {2\pi P_T\ov S} \) ^2   \ .
}
In particular, for a Schwarzschild black hole one simply has
$$
M^2_{\rm Rind}\equiv p_0^2-  p_1^2- p_\a^2 =
{4\hat N_L} \ .
$$ 
$M^2_{\rm Rind}$ represents the invariant mass in the Rindler Cartesian plane. 
 For a black hole
with $mw\neq 0$, $p_0,p_1$ are not conserved quantum numbers.

In the extremal limit
$\a ,\b , \g \to \infty $, $r_0\to 0$, with the charges \charges\ 
fixed, one obtains 
$$
A_{\rm ext}= B_{\rm ext}= {1\ov \sqrt{P}} \ ,\ \ \ \ \td \g_{\rm ext}
= {P_T \ov P}\ ,
\ \ \ \ {m\ov\td R}= w\td R \ ,
$$
where $P$ is the magnetic charge introduced earlier.
The Hamiltonian \mmma\ becomes
\eqn\hext{
\hat H_{\rm ext}  =
\hat N_R +\hat N_L +\ha p_\a^2 +{1\ov P} \[
\( {S\ov 2\pi}- P_T \)^2 - (P_R-P_L)\( {S\ov \pi} - P_T \) \] \ ,
}
$$
S=2\pi \sqrt{ mw P} \ .
$$
The Hamiltonian contains a free part and a  term
 multiplied by an overall factor 
$1/P$, which involves only the product of the three charges, i.e.
it depends only on the entropy and the Rindler energy operator.

\newsec{The  string source }

\subsec{Rindler energy }

For a classical black hole geometry, 
the energies measured by a Minkowskian distant observer (the ADM energy)
and by an inertial observer moving in the neighborhoods of the
horizon (where the $r,t$ part of the metric describes a
two-dimensional Rindler space-time) are  related by
\eqn\admm{
dE_{\rm R}={dE \ov 2\pi T_H}\ .
}
Using the first law of black hole thermodynamics, $dE=T_H dS$, we
can express the Rindler energy in terms of the entropy:
\eqn\fttt{
E_{\rm R}={S (E) \ov 2\pi } ={A\ov 8\pi G}\ .
} 
Consider for example the Reissner-N\"ordstrom black hole. From eq. \admm\
we have
\eqn\rnrnrn{
dE_{\rm R}={r_+^3\ov r_+^2-r_-^2}dE\ , \  \ \ \ r_+^2=\mu +\sqrt{\mu^2-Q^2}\ ,
\ \ \ \mu={4GE\ov 3\pi }\ .
}
Integrating this relation we obtain 
\eqn\eerr{
E_{\rm  R}={\pi\ov 4G} r_+^3\ .
}
In particular, for a neutral black hole, this gives 
\eqn\neutral{
E_{\rm R}={2\ov 3} E^{3/2} \( {8G\ov 3\pi} \) ^{1/2}\ .
}
This agrees with a  derivation given in ref. \ssk .

\subsec{Rindler energy of  states with $N_L=P_L=0$}

Equation \fttt\
can be explicitly checked for the black hole \solnfd \ by direct integration.
One obtains a relation of the form, $2\pi E_{\rm R}= S(E, m, w, P)$, which implicitly determines the ADM energy $E$ in terms of the Rindler energy and
the charges, i.e. $E=f(E_R,m,w,P)$. 

Let us now return to the physical spectrum of the conformal model.
Equation \spec \ can  be viewed as determining the eigenvalues of the Rindler operator
$P_T$ in terms of the charges and the oscillator state characterized by 
$\hat N_L, \hat N_R$ and $P_R-P_L$. 
The Rindler operator generates translations in the Rindler time $T$, and thus
its eigenvalues determine the Rindler energy of the configuration.

Consider the particular case of black holes with zero magnetic charge. Let us calculate the Rindler energy of  a state with 
$\hat N_L=0$, $P_L=0$. 
We will assume that $\hat N_R=mw\gg 1$, so that the Regge intercepts can be ignored, and consider the center-of-mass frame where
$p_\a ^2=0$. 
The Virasoro conditions 
\soro , \spec\ imply
\eqn\ssss{
\( {m\ov \td R}+ w\td R \) ^2 \( 1- {2\pi P_T\ov S} \) ^2 = 0 \ ,
}
or
\eqn\vals{
P_T={S\ov 2\pi } \ .
}
Strikingly, this is  the relation \fttt . 
Since these states carry the same charges $m,w$ of the ($P=0$) 
black hole, this means
 that the associated ADM mass is the black hole ADM mass!
Thus, for a black hole with zero magnetic charge,
an oscillating string in the horizon
in a $N_L=P_L=0$ state can be interpreted as a source of the gravitational field: it has the same charges and ADM mass of the black hole geometry.

\subsec{String sources for the non-extremal black hole 
with zero magnetic charge}

We have seen that the ADM mass eigenvalue associated with
states with only right-moving mode content
coincides with the ADM mass of the electrically charged black hole.
Let us now find the  general  states 
whose mass equals that of the $P=0$ black hole geometry.
To address this question, we return to the spectrum \spec\ and
look for states where the eigenvalue of the Rindler energy
operator is $P_T=S/2\pi $. For these states, there is a remarkable
simplification in the Hamiltonian \spec , which takes the form
\eqn\kkkk{
\hat H=\hat N_L +\hat N_R -{2\pi wm\ov S} (P_R-P_L) \ .
}
Using eq. \soro , and $S=2\pi (P_L+P_R)$, we obtain
\eqn\mmmaa{
P_R={S\ov 2\pi wm} N_R \ ,\ \ \ \ \ P_L=-{S\ov 2\pi wm} N_L\ .
}
All states in the physical spectrum satisfying \mmmaa\ will have
ADM mass eigenvalue coinciding with the ADM mass parameter
of the black hole configuration, and thus they can be used as
sources.
Note that the derivation leading to \mmmaa\ is not valid when $mw=0$, which includes the Schwarzschild geometry.
As pointed out above, in this case the zero-mode structure changes,
and there is an additional term in the Hamiltonian
of the form $-p_0^2+p_1^2$.

A natural question is  whether the logarithm of the total number $\Omega (E)$
of  states with ADM energy $E$ is  related to the Bekenstein-Hawking formula. 
The condition \mmmaa\ fixes the expectation value of the operators
$P_{R,L}$ in terms of $N_{R,L}$.
Ignoring power-like factors, the number of states obeying \mmmaa\ 
is asymptotically given by
\eqn\aaa{
\Omega (E)= \sum_{N_R=mw}^\infty 
 \exp \[ 2a_R \sqrt{N_R}+  2a_L \sqrt{N_R-mw } \] \ ,
}
where for the type II superstring theory, $a_{R,L}=\sqrt{2}\pi $.
The sum   over $N_R$    clearly diverges.  It is likely that
only a subset of the states obeying \mmmaa\ can constitute
the microstates of the black hole. 
This is the case for
the limiting situation of an extremal black hole, which is $N=1$ supersymmetric so it can only be identified with states with $N_L=0$. This fixes $N_R=mw$, so only one term in the sum
\aaa\ remains. 
The number of BPS states with the same asymptotic parameters of the 
black hole is $\Omega (E)\sim \exp [{\rm const.}\sqrt{ mw }]$.
However, the extremal black hole with $P=0$ is not suitable for a comparison
with the Bekenstein-Hawking prediction. As discussed in ref.~\sen ,
in this case $S\to 0$, the event horizon becomes singular, and the 
semiclassical calculation leading to the Bekenstein-Hawking formula does not 
apply. It would be interesting to understand if for the non-extremal
 black hole there is an extra condition restricting the set of states 
that contribute to the sum \aaa , which in the extremal limit reduces to $N_L=0$.

To summarize, the horizon region of a class of non-extremal black holes,
which includes the standard five-dimensional Reissner-N\"ordstrom solution, is
described by a solvable conformal field theory, and exact formulas
for  the physical spectrum can be written.
There is an ADM mass that can be associated with each
state of the  spectrum \spec . This is done by means of the eigenvalue of the 
Rindler energy operator $P_T$ on the state,
  using the standard relation between Rindler energy and ADM mass. 
We have seen that states  with Rindler energy  equal to the black hole
entropy over $2\pi $ are  special; in particular, the Hamiltonian in this sector
simplifies. 
To address counting problems, there are some points in the interpretation
of the spectrum that need to be clarified.
A thorough study is desirable to understand to what extent
these states can be regarded as  microstates constituting the 
non-extremal black hole.

\bigskip
\noindent {\bf Acknowledgements}: The author is grateful to A. Sen for
a useful discussion. He also wishes to thank G. Horowitz and
A. Tseytlin for comments.

\appendix {A}{Direct solution of the horizon model}
The general solution to the equations of motion corresponding 
to the Lagrangian \lagra\ is
given in terms of three free fields $Z, Y^+, Y^-$, as follows:
\eqn\soll{
z=Z -A \td \vp \ ,\ \ \ \  T= A (Z_R - Z_L ) 
+\ha \ln {Y^+\ov Y^-}\ , \ \ \ \td \rho^2= Y^+Y^-\ ,
}
$$
Z=Z_R(\s_-)+ Z_L(\s_+)\ ,\ \ \ \ Y^\pm=Y^\pm_R(\s_-) + Y^\pm _L(\s_+)\ ,
$$
$$
\del_\pm \td \vp =\pm \ha (Y^- \del _\pm Y^+ - Y^+ \del _\pm Y^-)\ ,
$$
so that
\eqn\axxx{
X^\pm=\td \rho \ \exp [\pm A(Z_R-Z_L)] \ Y^\pm \ .
}
We recall that superscripts $+$ and $-$ represent light-cone Lorentz indices, which are not in association with $\s_\pm $.
The boundary conditions \bbbb\ and \bbzz\  are satisfied by (throughout
we assume $mw\neq 0$):
\eqn\sooo{
Y^\pm _{R}=e^{\mp 2\td \g \s_- } \chi _{R}^\pm \ ,\ \ \ \ 
Y^\pm _{L}=e^{\pm 2\td \g \s_+ } \chi _{L}^\pm \ , 
}
$$
\td \g=\vep + Ap_z\ ,\ \ \ \ 
\chi_{R,L}^\pm (\s+\pi ) =\chi_{R,L}^\pm (\s )\ ,
$$
\eqn\yyyy{
Z_{L,R} = \s_\pm p_\pm + \hat Z_{L,R}\ ,\ \ \ \ \ 
\hat Z_{L,R}(\s+\pi )=\hat Z_{L,R}(\s )\ ,
}
$$
p_\pm = \pm (w \td R -A P_T ) + p_z \ ,\ \ \ \ \ P_T=P_R+P_L\ , 
$$
\eqn\aggg{
P_{R,L}={1\ov 4\pi } \int _0^\pi d\s
(Y^- _{R,L}\del _\mp Y^+_{R,L} - Y^+_{R,L} \del _\mp Y^- _{R,L} )\ .
} 
The Fourier expansions $\chi ^\pm_{R,L} $ and $\hat Z_{R,L} $ are
given by 
\eqn\four{
\chi^\pm _R= {x^\pm \ov\sqrt{\td \g}} + {i \ov \sqrt {2}}\sum_{n\neq 0} 
x^\pm_n e^{-2in\s _- }\ ,\ \ \ \ \chi^\pm _L= 
{\td x^\pm \ov \sqrt{\td\g } }+ {i \ov \sqrt {2}}\sum_{n\neq 0} 
\td x^\pm_n e^{-2in\s _+ }\ ,}
\eqn\fourier{
\hat Z_{R}=z_0+
{i \ov \sqrt {2}}\sum_{n\neq 0} {\a _{nz}\ov n} e^{-2in\s _- }\ , \ \ \ \ 
Z_{L}=\td z_0+
{i \ov \sqrt {2}}\sum_{n\neq 0} {\td \a _{nz}\ov n} e^{-2in\s _+ }\ .
}
Starting from the expression for the classical stress-energy
tensor of the theory \lagra\ and evaluating it on the general solution
\soll , one finds that it takes the simple form:
\eqn\ttmmm{
T_{\pm\pm} =\del_\pm Z_{L,R} \del_\pm Z_{L,R} + \del_\pm Y^+ \del_\pm Y^- \ .
}
The canonical commutation relations for $X^\pm $ imply
\eqn\canon{
[ x^+_n, x^-_k]= 2(n + i\td \g )^{-1}\delta _{n+k} \ ,\ \ \ \ 
[ \td x^+_n, \td x^-_k]= 2(n - i\td\g )^{-1}\delta _{n+k}\ , 
}
$$
[x^-,x^+]=i=[\td x^+, \td x^-] \ .
$$
In the Euclidean theory the ``$i$" is absent from eq. \canon , and one can
introduce the creation and annihilation operators $b_{n\pm}, b_{n\pm}^\dagger$
of eqs. \nnn , \angulr\ by rescaling the modes by factors $(n\pm \td \g)^{1/2}$.
In the Minkowskian theory one may formally define
$b_{n\pm}, b_{n\pm}^\dagger$ operators by rescaling by factors 
$(n\pm i\td \g)^{1/2}$, but the $b^\dagger_n$ are no longer the Hermitian conjugate of the
$b_n$; the notation becomes misleading, 
so it is convenient to replace them by  the standard $\a_n^\mu $ operators
as in eq.~\alfa .

The light-cone gauge can be fixed by setting to zero 
the oscillator part of $\chi ^+_{R,L}$ ; as usual, the oscillators
of $\chi ^-_{R,L} $ are then determined in terms of transverse oscillators
by imposing the Virasoro constraints $L_n = \td L_n=0 \ ,\ n\neq 0$.
After a bit of algebra one obtains the quantum Virasoro operators, 
$\hat L_0={1\ov 4\pi }\int _0^\pi d\s T_{--}$, 
$\hat {\td  L_0}={1\ov 4\pi }\int _0^\pi d\s T_{++}$, as given by
eqs.~\mmma , \soro , with the $P_{R,L}$ and $ N^{(R)}_R$ of eqs.~\pppr , 
\otro , and similar expressions for $N_L^{(R)}$. In the NS sector
the expressions are also similar, with the usual changes 
that take place in  free superstring theory.

\vfill\eject
\listrefs
\end